## Abstract

As an on-ramp to databases, we offer several well-structured private database templates as open-source resources for agriculturalists – particularly those with modest spreadsheet skills. These farmer-oriented Airtable databases use simple data-validated forms, with the look and feel of a customized app, to yield operational data that is tidy, machine- and human-readable, editable, and exportable for analysis in other software. Such data can facilitate logistics, provide contextual metadata, and improve enterprise analysis. A recorded workshop explaining how to build a database for activity records is presented. These resources may facilitate infusion of digital agriculture principles through Extension and structured educational programming.

## Introduction

Farmers' commitment and ability to keep good records varies tremendously. The idea of records and notes, often done in notebooks, dates to at least the mid-1800s (Joly, 2011). Unfortunately, these notes are often cryptic, misplaced, or damaged and for many, remain unused. If such information were recorded digitally and stored in the cloud, we immediately solve some access and consistency issues and make this data FAIR (*findable*, *accessible*, *interoperable*, *reusable*). More importantly, interoperable digital formats can also enable mining for insights and analysis in simplistic (sorting or filtering) or complex (multifactor analysis on yields, harvest date projections, etc.) manners.

While farmers and agricultural advisers frequently use spreadsheets like those offered thru the tremendous library offered by public and land grant institutions. databases are less commonly used by them. A weak link is knowing inputs to such tools. Farm management information systems (FMIS), original equipment manufacturer (OEM) platforms, and other cloud services are making strides toward interoperability (AgGateway, 2021) to provide this data, but gaps remain as detailed information that only a human knows may be overlooked. A database populated at the time of an operation could produce high quality data for those decision aids.

Although spreadsheets can be set up as databases, database software has capabilities that increases the capacity to synthesize data. Google Forms, for example, can be used for data entry into a Google Sheet (Hickson, 2020) and will work in some instances, but requires pseudo-coding and planning to make sure the data is ultimately usable and complete. Airtable (Meijer, 2022; https://airtable.com/) offers similar capacities with the benefit of improved data validation for form completion and the look and feel of customized apps.

Database applications are well beyond activity records; but our focus here is on activity records because of their importance in making logistical and strategic decisions, and their ability to provide the fuller contextual metadata to drive artificial intelligence and machine learning algorithms or support economic analyses. The content of this article can address the 11 types of farm records that Apindi and Simwa (2022) recommend keeping. With the overall goal of providing a simple primer to databases and a logical on-ramp for famers and advisers to digitize records effectively, our specific objectives for this article are to:

1. Explain basic database features so Extension educators and trainers can extend this knowledge.
2. Illustrate the value of well-structured tidy and open data (i.e., simplicity and interoperability with spreadsheets or other software).
3. Introduce several prepared Airtable templates and related resources.

**Database basics**

Spreadsheets such as Microsoft Excel or Google Sheets are organized as workbooks with sheets which can be renamed, cross-referenced, etc. Airtable looks and feels much like a spreadsheet, but it is a base with interactive tables. This makes it a good entry point for novices. It is platform-agnostic and can be used in-browser or as a native app on Android or iOS devices. Airtable has a free tier that should suffice in most farm situations.

Airtable is cloud-based. It can be used offline, however, online usage brings practical benefits of real-time saving and sharing. "Sharing" has assorted levels of security with permission levels of *owner/creator*, *editor*, *commenter*, or *read-only*. Data entry forms can be completed by workers or affiliates who cannot access the database itself.

The templates introduced below employ data validation. The preset lists of options (such as names of employees or equipment) enable users to choose from drop-downs list rather than type freeform values. Benefits are fewer errors, improved consistency, and less tedium (being able to choose "24 row planter" from the list rather than typing it out); with automatic sorting, the most recently used option can appear at the top of the list the next time the form is completed. The templates also apply conditional data requests to simplify the forms; only data pertinent to the operation is requested.

**Using well-structured tidy data**

Data in Airtable tables will be organized as tabulated tidy data (Neo, 2020). "Tidy" means each column is a different variable, each row is an observation, and each cell is an individual value. Table 1. Illustrates tidy data from the Horticultural Crop Activity Records (Buckmaster, 2023a). Values can be of different types such as integers, real numbers, text, hyperlinks, or documents (e.g., photo or other files). Tidy data facilitates sorting, filtering, and interpretation. It also facilitates export and usage in other software (such as exporting to a spreadsheet to do pivot tables).

**Templates already generated**

Wiginton (2022) identified excellent Airtable examples and use cases as diverse as accounting, customer relationship management, marketing, fitness tracking, timesheets, and live dashboards.

Table 1. Sample subset of tabular data (grid view) exported from the horticultural crop activity records database template as CSV into Microsoft Excel.

| Who | Where | What | Duration | Notes | created time | Power Unit | Implement(s) | Seeds planted | Seeding Rate (seeds/ac) | Products applied | Fertilizers applied | Fertilizer Rate (lb/ac) |
|---|---|---|---|---|---|---|---|---|---|---|---|---|
| Purdue Pete | Bed 72 | Tillage | 40 | left disc needs adjustment | 12/20/2022 11:35am | Tractor 2 JD X120 | bed shaper | | | | | |
| Suzie Jones | Bed 72 | Plant/Transplant | | | 12/20/2022 11:37am | Utility tractor | water wheel transplanter | onions - candy | | | | |
| Suzie Jones | Bed 72 | Harvest | 30 | bundles of rhubarb | 12/20/2022 11:50am | human powered | | | | | | |
| Purdue Pete | Field 1 | Spread/Spray | 120 | | 12/20/2022 11:39am | Gator | 150 gal sprayer | | | Glyphosate | | |
| Suzie Jones | Field 1 | Plant/Transplant | | burn down looked effective | 12/20/2022 11:41am | Tractor 2 JD X120 | seed planter | corn - sweet - 82 day | 30000 | | 9-18-9 starter | 50 |
| Purdue Pete | Field 1 | Scout | | popcorn is near ready | 12/20/2022 11:42am | | | | | | | |
| Purdue Pete | Zone D | Scout | | all looks great | 12/20/2022 12:30pm | | | | | | | |

Specific to production agriculture, Digital Field Records (Buckmaster 2022) is an Airtable template released with video documentation and brief tutorials that enable farmers or researchers to readily collect digital records of "what happened here?" for wide-acre field and forage crop operations. Data validation simplify data entry (Figure 1) but options for entering free-form text or uploading photos remain. Users can quickly duplicate this free template and customize it (including permissions) for their own operation. The records from this private database would be of value in tracking season progress, documenting machinery usage and labor efficiency, supporting claims about the crops produced (e.g., organic, GMO free), and informing enterprise budgets. These records can also round out information from FMIS or OEM platforms with necessary data regarding climate and carbon balance (Illinois Soybean Association, 2022).

A derivative customized for Horticultural Crop Activity Records (Buckmaster, 2023a) – also with tutorial videos – accommodates specialty crop operations of all types of areas as large as fields or in diverse production areas such as high tunnels, benches, zones, blocks, or beds. The resulting data could inform community supported agriculture (CSA) schedules and inventories and improve seeding and transplanting management in coming years.

A Digital Food Safety Modernization Act (FSMA) toolkit (Buckmaster et al., 2022) was also designed using Airtable; its purpose is to make FSMA records easier and more complete for those who are unfamiliar with FSMA requirements or for whom it is simply tedious or confusing. Using this toolkit will ensure compliance with FSMA. The Digital Marketing and Delivery records template (Buckmaster and Soonthornsima, 2022) is useful for tracking contracts and deliveries of commodities in simple tables that can facilitate reconciliation and informed decision-making. With such records completed as each contract or delivery is made, users will have fingertip access to information about unmet commitments and can evaluate marketing decisions with regard to futures price risk and basis risk separately.

These templates and associated video series offer immediate value to users and can spawn ideas of other important records of events, activities, or decisions to digitize. Simple Personal Databases (Buckmaster, 2023b) shares five brief instructional videos that explain how to generate and use a customized activity record database in Airtable.

Figure 1. Sample Airtable form from the digital field records template.

**Conclusion**

Private databases can be simple and useful introductory tools for farmers and others wanting to digitize activity records. This approach uses simple forms completed with data validation which simplifies data input and makes it more consistent. Data is updated in real-time. Several Airtable database examples were introduced with pointers to online resources that fully document how to duplicate and customize each reference implementation. These templates are practical for production and research farms and serve as an introduction to more complex databases and analysis for applications well beyond production agriculture.

Buckmaster, D.R., M. Basir, and H. Sakata. 202x. **Facilitating Digital Agriculture with Simple Databases.** *Journal of Extension (Tools of the Trade | in press*).